\documentclass[aps,prl,twocolumn,superscriptaddress,showpacs]{revtex4}

\usepackage[dvips]{graphicx}
\usepackage{latexsym}

\begin{document}

\title{Anisotropic diamagnetic response in type-II superconductors 
with gap and Fermi surface anisotropies
}

\author{
H. Adachi
}
\email[]{adachi@mp.okayama-u.ac.jp}
\affiliation{
Department of Physics, Okayama University, Okayama 700-8530, Japan
}
\author{
P. Miranovi\'{c}
}
\affiliation{
Department of Physics, University of Montenegro, Podgorica 81000, Serbia and Montenegro
}
\author{
M. Ichioka
}
\affiliation{
Department of Physics, Okayama University, Okayama 700-8530, Japan
}
\author{
K. Machida
}
\affiliation{
Department of Physics, Okayama University, Okayama 700-8530, Japan
}

\date{\today}

\begin{abstract}
Effects of anisotropic gap structures on a diamagnetic response 
are investigated in order to demonstrate that 
the field-angle-resolved magnetization ($M_L(\chi)$) 
measurement can be used as a spectroscopic method to detect gap structures. 
Our microscopic calculation based on the quasiclassical Eilenberger formalism 
reveals that $M_L(\chi)$ in a superconductor with four-fold gap displays 
a four-fold oscillation reflecting the gap and Fermi surface (FS) anisotropies, 
and the sign of this oscillation changes at a field between $H_{c1}$ and $H_{c2}$. 
As a prototype of unconventional superconductors, 
magnetization data for borocarbides are also discussed. 

\end{abstract}


\maketitle
Precise determination of the node position or gap structure is of 
fundamental importance in superconductivity study in general, 
especially for ever-growing so-called unconventional superconductors 
since it is indispensable in identifying the pairing mechanism for 
a material of interest. 
There are only a few established methods for the precise determination of 
gap structures; 
Angle-resolved specific heat and thermal conductivity 
measurements are notable ones~\cite{Izawa, Park1, Izawa2, Aoki, Pedja}. 
Nevertheless, to reinforce the conclusion, 
more such spectroscopic experiments 
based on bulk quantities are desirable~\cite{motiv}.
As Takanaka~\cite{Takanaka} suggested within the Ginzburg-Landau (GL) theory, 
a diamagnetic response is 
a strong candidate if combined with an analysis of the basal plane 
magnetization anisotropies. 
Needless to say a diamagnetic response from a superconductor is a hallmark of 
rigidity of the macroscopic wave function, 
containing a wealth of microscopic information, 
and it is a routine work to measure magnetizations to check 
if a material of interest is a superconductor or not. 
Since we are realizing~\cite{Pedja} 
that field-angle-dependences of various physical quantities 
such as specific heat or thermal conductivity should reflect 
low-lying quasi-particles around the vortex core, 
it is also expected that the magnetization contains the same kind of information. 

Among a vast amount of type-II superconductors 
nonmagnetic borocarbides $R$Ni$_2$B$_2$C ($R$=Lu,Y) 
are considered to be typical examples of unconventional ones
in the following sense: 
(i) Lots of experiments have demonstrated the existence of gap nodes (or gap minima)
~\cite{Izawa, Park1, Nohara, Yokoya, Yang}, 
(ii) the normal and mixed state 
are not exposed to the strong fluctuations 
such as in cuprates, 
(iii) there exist detailed magnetization measurements~\cite{Kogan1,Civale1}.
To establish a spectroscopic method based on the magnetization measurements 
it is thus important to understand 
the anisotropic diamagnetic response in these materials 
and clarify its relation to the gap structure. 

\begin{figure}[t] 
\scalebox{0.35}[0.35]{\includegraphics{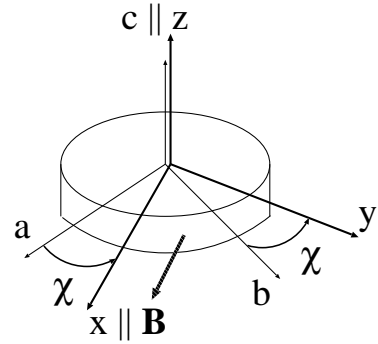}}
\caption{The coordinates $(x,y,z)$ and the crystal axes $(a,b,c)$.
The induction ${\bf B}$ is rotated from the $a$-axis by an angle $\chi$.
}
\label{fig:coordinate}
\end{figure}

In the works by Civale {\it et al.}~\cite{Civale1} and Kogan {\it et al.}~\cite{Kogan1}, 
basal plane magnetizations were measured as a function of the angle $\chi$ 
(see Fig. \ref{fig:coordinate}) 
between the applied field and the crystal axis, 
and it was found that the four-fold oscillation of the magnetization showed a 
sign reversal with decreasing the field (or temperature). 
Kogan {\it et al.}~\cite{Kogan1} demonstrated that these behaviors can be 
reproduced within a nonlocal London theory 
without quoting anisotropy effects of 
the gap function.
 From a theoretical viewpoint it is quite reasonable to consider that 
the lower field (or lower temperature) behavior is well described within this framework 
since the London theory is believed to be appropriate for studying a vortex state in 
a field region $H \ll H_{c2}$. 
 With increasing the field, however, the vortices start to overlap and 
the physics of vortex cores begins to play an essential role, 
whereas the core physics cannot be captured 
by any London theory without introducing some cutoff function by hand.
 As a consequence, at least when discussing anisotropy effects in a vortex state, 
the validity of London description in high fields is quite unclear. 
If we aim to clarify whether the observed phenomena are generic ones or not, we need a theoretical approach which can correctly describe the anisotropy of $H_{c2}$ and the core effects. 

The purpose of this Letter is two-fold. 
The first one is to clarify the effects of gap structures 
on the anisotropic diamagnetic response based on the quasiclassical 
Eilenberger formalism. 
The second one is to apply our analysis to 
the prototype materials, i.e., nonmagnetic borocarbides, and 
present a microscopic description of the observed mysterious sign reversal of 
$M_L(\chi)$-oscillation~\cite{Civale1,Kogan1}. 
Our microscopic treatment covers, of course, both GL theory and 
London theory, since the former is derived from the Eilenberger formalism 
through an expansion about the pair field, while the latter by using 
a phase only (London) approximation. 
Indeed our numerical solution at finite temperatures well reproduces 
both GL behavior $( M \propto H_{c2}-B )$ near $H_{c2}$ and London behavior 
$(M \propto \ln (H_{c2}/B) )$ in lower fields. 

Let us first explain our numerical procedure. 
We start with the following Eilenberger equation~\cite{Eilenberger}
($k_{\rm B}=\hbar=1$)
\begin{equation}
  f( {\varepsilon_n}, {\bf p}, {\bf r})
=  
(2 \varepsilon_n+ {\rm i} {\bf v \cdot \Pi} )^{-1} 
\Big( 2
 g( {\varepsilon_n}, {\bf p},{\bf r}) \, w_{\bf p} \Delta({\bf r})
\Big), 
\label{eq:Eilen1} 
\end{equation}
where $f({\varepsilon_n},{\bf p},{\bf r})$, 
$f^{\dagger}({\varepsilon_n},{\bf p},{\bf r})= 
f^*({\varepsilon_n},-{\bf p},{\bf r})$, and 
$g({\varepsilon_n},{\bf p},{\bf r}) = 
\sqrt{1- f({\varepsilon_n},{\bf p},{\bf r})f^{\dagger}({\varepsilon_n},{\bf p},{\bf r})}$
are the Eilenberger's Green's functions. 
Here $\varepsilon_n= 2 \pi T(n+1/2)$ is a fermionic Matsubara frequency, 
${\bf v}$ is a Fermi velocity, 
${\bf \Pi}=-{\rm i} {\bf \nabla}+ (2 \pi/ \Phi_0){\bf A}$ is a 
gauge invariant gradient, $T_c$ is a transition temperature at a zero-field, 
and $\Phi_0$ is the flux quantum. 
The gap function is expressed as 
$\Delta_{\bf p}({\bf r})= w_{\bf p} \Delta({\bf r})$ where 
$w_{\bf p}$ is the pairing function with relative momentum ${\bf p}$ of 
the Cooper pair, 
and $\Delta({\bf r})$ is the order parameter with center of mass coordinate ${\bf r}$. 
Throughout this paper 
we treat extreme type-II superconductors with large GL parameter $\kappa \gg 1$ 
(in borocarbide superconductors $\kappa \!\agt\! 10$), 
so that the vector potential 
is approximated by ${\bf A}= -Bz {\bf \hat{y}}$ 
where ${\bf B}=B {\bf \hat{x}}$ is the induction (see Fig. \ref{fig:coordinate}).
This is indeed a good approximation in high-$\kappa$ case 
since the correction term is of order $O(1/\kappa^2)$. 
To solve Eq. (\ref{eq:Eilen1}), we adopt an approximation similar 
to that used by Pesch~\cite{Pesch}, 
\begin{equation}
  f \approx 2g w_{\bf p} 
\left( 2\varepsilon_n + {\rm i}{\bf v \cdot \Pi} \right)^{-1}\Delta. 
\label{eq:Pesch}
\end{equation}
The physics behind this approximation is that the spatial variation of $f$ related to 
the phase modulation of $\Delta$ is much larger than the spatial variation of $g$ 
describing the amplitude fluctuation. 
It is worth noting that we do not replace $g$ in the above equation 
by its spatial average as Pesch done, 
in order to ensure that the correct expression for the nonlocal GL free energy $F/V$ 
given in Ref. \onlinecite{Adachi} are reproduced up to the quartic term. 
Besides the above mentioned justification near $H_{c2}$, the applicability to lower 
fields is improved by requiring the selfconsistency among $f$, $f^{\dag}$ and $g$. 
Furthermore our scheme can be valid in an anisotropic case by including 
the contribution of higher Landau level components to $\Delta$: 
\begin{eqnarray}
  \Delta &=& \Delta_0 \sum_{N =0}^{N_{\rm max}} d_N \psi_N, \label{eq:LLX1}\\
  \psi_N &=& 
  \sum_{m=-\infty}^{\infty} C_m 
  \frac{ H_N(z+\nu m)}{\sqrt{2^N N!}} e^{-(z+\nu m)^2/2-{\rm i}\nu m y }, 
 \label{eq:LLX2}
\end{eqnarray}
where $\Delta_0=1.764 T_c$, $r_B= \sqrt{\phi_0/2 \pi B}$, 
$C_m= ( {\nu r_B}/{\sqrt{\pi}} ) e^{- {\rm i} \pi \zeta m^2}$, 
$H_N$ is the $N$-th Hermite polynomial, and the lengths are measured 
in units of $r_B$. 
The real constants $\zeta$ and $\nu$ specify the configuration of a vortex lattice. 
Since the difference of a vortex lattice configuration is considered to be irrelevant 
to the quantity in question, we set in this paper $\zeta=1/2$ and 
$\nu= \sqrt{\sqrt{3} \pi}$; the value for a triangular lattice. 
Substituting the above expression into Eq. (\ref{eq:Pesch}) and using a 
parameter representation 
\begin{equation}
\left( 2\varepsilon_n + {\rm i}{\bf v \cdot \Pi} \right)^{-1}
=
\int_0^{\infty} d \rho e^{-( 2\varepsilon_n + {\rm i}{\bf v \cdot \Pi} ) \rho}, 
\end{equation}
we have 
\begin{equation}
  f= 2 g w_{\bf p} \int_0^{\infty} d \rho e^{-2 \varepsilon_n \rho}
  \Big( \Delta_0 \sum_{N =0}^{N_{\rm max}} \alpha_N d_N \Big). 
\label{eq:Eilen2}
\end{equation}
Here the expression for $\alpha_N \equiv e^{-{\rm i}\rho {\bf v \cdot \Pi}} \psi_N$ 
is given by  
\begin{eqnarray}
  \alpha_N 
  &=& 
  \sum_m C_m
  \frac{ H_N(z+\nu m- {\rm Re}\, \lambda)}{\sqrt{2^N N!}} \nonumber \\
  && \qquad 
\times e^{-(|\lambda|^2- \lambda^2)/4}e^{-(z+\nu m- \lambda)^2/2-{\rm i}\nu m y }, 
\end{eqnarray}
where $\lambda= (v_{z}+ {\rm i}v_{y})\rho/r_B$. 
At long last we have a solution for $f$, 
on condition that we have the correct $\{ d_N \}$-values. 
To determine $\{ d_N \}$ we use 
the following selfconsistent equation projected onto each Landau level:
\begin{equation}
  \Big( \ln (\frac{T}{T_{c}}) 
+ 2 \pi T \sum_{n \ge 0} \varepsilon_n^{-1} \Big) d_N 
=
   \frac{2 \pi T}{\Delta_0} \sum_{n \ge 0}
  \overline{ \psi_N^* \langle  w_{\bf p}^* f \rangle}, \label{eq:ite1} 
\end{equation}
where the overbar denotes the spatial average, and 
the FS average $\langle \cdots \rangle$ satisfies a normalization 
condition $\langle 1 \rangle=1$; the definition is given by Eq. (\ref{eq:FSave}) below. 
Our numerical procedure is as follows:
Input initial values for $\{ d_N \}$, $f$, $f^\dagger$ and $g$. 
Next use Eq. (\ref{eq:Eilen2}) to obtain the new $f$ (and $f^\dagger$, $g$). 
Then iterate Eq. (\ref{eq:ite1}) to renew the $\{ d_N \}$- values, 
and return to Eq. (\ref{eq:Eilen2}). 
In order to check the reliability of our numerical procedure, we initially 
treated a two-dimensional case and calculated field dependences of 
each $\{ d_N \}$ at $T/T_c=0.5$. 
The obtained result was quite similar to the previous work 
based on the Landau level expansion of the GL equation 
(Figure 2(b) of Ref. \onlinecite{Kita}). 
In the following calculation we use $N_{\rm max}=6$. 

\begin{figure}[t]
\scalebox{0.50}[0.50]{\includegraphics{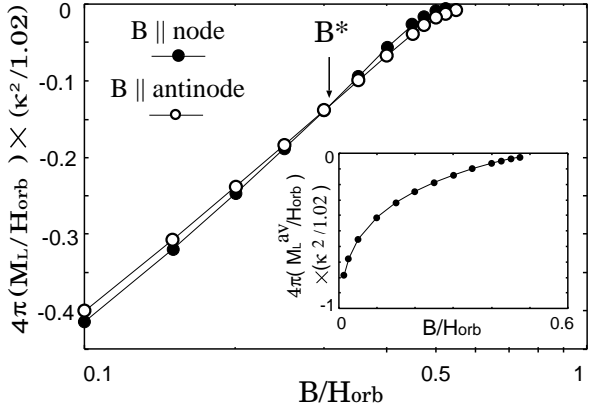}}
\caption{
Logarithmic field dependence of longitudinal magnetization $M_L$ 
at $T/T_c=0.6$ for ${\bf B} \parallel$ node (filled circles) and 
${\bf B}$ $\parallel$ antinode (open circles). 
The used anisotropy parameters are $\alpha=1$ and $\beta=0$. 
$H_{\rm orb}=1.037 \Phi_0/2 \pi \xi_0^2$
$(\xi_0= v_0/2 \pi T_c)$
is the orbital limiting field in the isotropic case. 
Inset: Field dependence of 
$M_L^{\rm av}=(M_L (\chi=0)+M_L(\chi=\pi/4))/2$ 
for the same parameters. 
}
\label{fig:magL_t06_a1b0}
\end{figure}
 
 The magnetization $4 \pi {\bf M}={\bf B}- {\bf H}$ is obtained from the relation 
${\bf H}= 4 \pi {\bf \nabla}_{\bf B} (F/V)$.
As for the longitudinal component ${\bf M}_{L} (\parallel {\bf H})$, 
which we focus on in this paper,  
Klein {\it et al.}~\cite{Klein} obtained a more convenient formula extending the 
virial theorem derived by Doria {\it et al.}~\cite{Doria}: 
\begin{eqnarray}
 -4 \pi M_{L}
&=&
 \frac{2 \pi^2 N(0)}{B} 
T \sum_{n \ge 0} 
\Big\langle 
\overline{
\frac{2g(f^{\dagger}(w_{\bf p}\Delta)+ f (w_{\bf p}\Delta)^*) }{1+g}
} \nonumber \\
&&  \hspace{3cm} - 4 \varepsilon_n \overline{(1-g)}   \Big\rangle,
\end{eqnarray} 
where in the above equation 
we approximately set ${\bf B \cdot H} \simeq BH$ as in Ref. \onlinecite{Kogan1}.

\begin{figure}[t]
\scalebox{0.50}[0.50]{\includegraphics{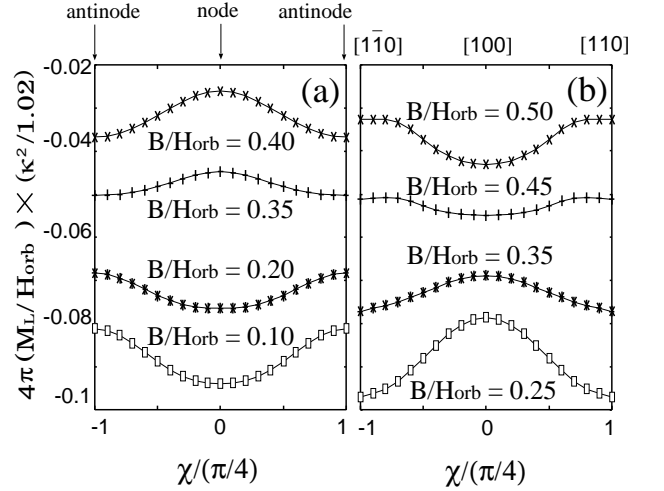}}
\caption{Field-angle dependences of $M_L(\chi)$ for (a) $\alpha=1, \beta=0$ 
and (b) $\alpha=1, \beta=0.4$ at $T/T_c=0.6$ for several inductions. 
The data at different $B$ are vertically shifted.
}
\label{fig:magL-chi_t06}
\end{figure}

 Now let us assume an isotropic FS to focus only on 
the role of the gap anisotropy. 
In this case 
the Fermi momentum can be written as 
${\bf p}= m v_0 \hat{r}$, where 
$\hat{r}= (\sin\theta \cos\phi, \sin\theta \sin\phi, \cos\theta)$, 
$v_0$ is the Fermi velocity in the isotropic case, 
and $m$ is the effective mass of the quasi-particle. 
As a model for an anisotropic gap function, 
$w_{\bf p}= \sqrt{1- \alpha \cos 4(\phi+\chi)}/
\langle 1- \alpha \cos 4(\phi+\chi)  \rangle$
is used where $\chi$ is the field-angle measured from the crystal $a$-axis, 
and $\phi$ is the azimuthal angle measured from the $x$-axis. 
Thus $\alpha$ denotes the degree of the gap anisotropy with $\alpha=1$ being 
the nodal case. 
Figure \ref{fig:magL_t06_a1b0} shows the field dependence of 
the longitudinal magnetization $M_L$ for ${\bf B} \parallel$ node (filled circles) and 
${\bf B} \parallel$ antinode (open circles) at $T=0.6T_{c}$. 
From the inset both the GL and the London behaviors are clearly seen. 
Although the difference of $M_L$ between the two field-orientations is rather small, 
we can find that the $M_L$ at $B \simeq H_{c2}$ is larger 
for ${\bf B}$ $\parallel$ node, but with lowering the field this tendency 
is reversed at a field $B^*$. 
In Fig. \ref{fig:magL-chi_t06}(a), the corresponding field-angle dependences 
of $M_L(\chi)$ are plotted for several inductions. 
It should be emphasized here that the gap anisotropy {\it alone} can cause a sign 
reversal of the $M_L(\chi)$-oscillation. 
The result with isotropic FS can be summarized as follows:
$M_L^{B \parallel {\rm node}} > M_L^{B \parallel {\rm antinode}}$ in higher fields, and
$M_L^{B \parallel {\rm antinode}} > M_L^{B \parallel {\rm node}}$ in lower fields.

Next we discuss the magnetization experiments~\cite{Civale1,Kogan1} for borocarbides. 
The observed data are incompatible with the above conclusion 
once we recall the experimental suggestion that the nodes 
exist along $[100]$ and $[010]$ directions~\cite{Izawa, Park1}. 
The discrepancy is considered to stem from the {\it unusually} large FS anisotropy 
possessing partial nesting~\cite{Dugdale} in these materials.  
As a model having such an anisotropic FS, 
we introduce a four-fold anisotropic dispersion 
\begin{equation}
  \epsilon_{\bf p}= \frac{1}{2m}\Big( (p_x^2+p_y^2)
(1+ \beta \cos 4(\phi+ \chi) ) + p_z^2 \Big) 
= 
\frac{p^2}{2m} \sigma^2, 
\end{equation}
where $\sigma(\theta, \phi)= \sqrt{1+ \beta \sin^2 \theta \cos 4(\phi+\chi) }$. 
Thus $\beta$ denotes the degree of the FS anisotropy. 
The resultant Fermi momentum can be written as 
${\bf p}= (m v_0/\sigma )\hat{r}$. 
The Fermi velocity ${\bf v}= {\bf \nabla}_{\bf p} \epsilon_{\bf p}$ 
can be expressed as 
${\bf v}= v_r \hat{r}+ v_{\theta} \hat{\theta} + v_{\phi} \hat{\phi}$, 
where $\hat{\theta}=(\cos\theta \cos\phi, \cos\theta \sin\phi, -\cos\theta)$ 
and 
$\hat{\phi}=(-\sin\phi, \cos\phi, 0)$.
Here each component of ${\bf v}$ is given by 
$v_r= v_0 \sigma$, $v_\theta=2 v_0 (\beta/\sigma) 
\sin^3\theta \cos \theta \cos4(\phi+\chi)$, 
and $v_\phi=- 2 v_0 (\beta/\sigma) \sin^3\theta \sin 4(\phi+\chi)$.
Finally, the area element $dS$ of the FS divided by $|{\bf v}|$ is given by 
$dS/|{\bf v}|= (m^2 v_0/\sigma^3)d(\cos\theta)d \phi$. 
Then our definition of the FS average is given by 
\begin{equation}
  \langle A_{\bf p} \rangle = 
\frac{ \int_{\rm FS} (dS \cdot A_{\bf p}/|{\bf v}|)  } 
{\int_{\rm FS} (dS/|{\bf v}|)}. 
\label{eq:FSave}
\end{equation}
A band structure calculation for LuNi$_2$B$_2$C suggests a rough estimate 
$\beta \simeq 0.4$ so as to reproduce the ratio 
$\langle v_a^4 \rangle/ \langle v_a^2 v_b^2 \rangle= 0.128$~\cite{Kogan3} 
within our model. 

 Starting from the isotropic FS case ($\beta=0$) and 
increasing the $\beta$-value, 
the oscillation pattern seen in Fig. \ref{fig:magL-chi_t06}(a)
first tends to diminish, 
and when the $\beta$-value exceeds about 0.1, the sign of the $M_L(\chi)$-oscillation 
pattern is completely reversed. 
This is shown in Fig. \ref{fig:magL-chi_t06}(b), and the oscillation behavior 
well coincides with the results of Refs. \onlinecite{Civale1} and \onlinecite{Kogan1}.
Worth noting is that $\alpha \beta>0$ in our model corresponds to 
the {\it competing anisotropy} case in the sense of Ref. \onlinecite{Nakai}, 
where the observed configuration~\cite{Eskildsen} of the vortex lattice in 
${\bf B} \parallel c$ is properly explained 
by the competition between gap and FS anisotropies. 
Note also that the main conclusion here is not changed 
by the nodal topology and effective dimensionality of a material, 
though the oscillation amplitude is {\it quantitatively} enhanced for a quasi 
two-dimensional material. 
\begin{figure}[t]
\scalebox{0.45}[0.45]{\includegraphics{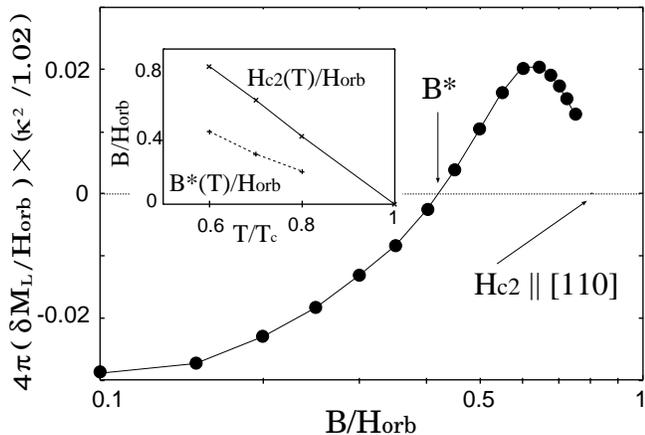}}
\caption{
Logarithmic field dependence of $\delta M_L= M_L(\chi=\pi/4)- M_L(\chi=0)$ 
for $\alpha=1, \beta=0.4$ at $T/T_c=0.6$. 
Inset: The sign reversal field $B^*$ in the $B$-$T$ phase diagram. 
The $H_{c2}$ is determined through linear extrapolations of $M_L$. 
} 
\label{fig:DmagL-hc2_a1b04}
\end{figure}

We show in Fig. \ref{fig:DmagL-hc2_a1b04} 
the field dependence of the magnetization oscillation amplitude 
$\delta M_L= M_L(\chi=\pi/4) -M_L(\chi=0)$ at $T/T_c=0.6$, 
which is to be compared with the experiments (Fig. 3 of Ref. \onlinecite{Civale1} 
and Fig. 5 of Ref. \onlinecite{Thompson}).
The characteristic behavior that $\delta M_L$ is a slowly increasing function 
of the field except for a peak structure slightly below $H_{c2}$ is well 
reproduced. 
Finally the inset of Fig. \ref{fig:DmagL-hc2_a1b04} shows 
the oscillation sign reversal field $B^*$ 
in the $B$-$T$ phase diagram for $\alpha=1$ and $\beta=0.4$.
The $B^*(T)$ is a decreasing function of the temperature, 
and this is consistent with the experimental finding of Ref. \onlinecite{Civale2}. 
The $B^*(T)$ could be a diagnostic quantity to characterize the gap function 
and the FS anisotropy. 

The physics behind the phenomena is explained as follows:
Near $H_{c2}$ where the GL theory can be applied, 
the anisotropy of the longitudinal magnetization 
$ 4 \pi M_L \simeq (B-H_{c2})/2.32 \kappa^2$ 
comes from that of $H_{c2}$. 
In lower fields where the London theory is more appropriate, 
the anisotropy of 
$4 \pi M_L \simeq (-\Phi_0/8 \pi \lambda^2) \ln (H_{c2}/B)$ 
is {\it effectively} attributed to that of the penetration depth $\lambda$. 
Extending a knowledge on the relation between 
the $\xi$- and $\lambda$-anisotropies 
based on the anisotropic GL equation~\cite{Gorkov}, 
a naive relation 
$H_{c2}(\chi=0)/H_{c2}(\chi=\pi/4) \sim \lambda^2(\chi=0)/\lambda^2(\chi=\pi/4)$
is expected to hold, where $\lambda(\chi)$ means an {\it effective} penetration 
depth {\it perpendicular} to $H_{c2}(\chi)$. 
Namely if $H_{c2}$ is larger, then $1/\lambda^2$ is expected to be smaller, and 
this causes the sign reversal of the $M_L(\chi)$-oscillation. 

Before ending we briefly discuss the origin of the peculiar angular variation 
of $M_L(\chi)$ seen in Refs. \onlinecite{Civale1} and \onlinecite{Kogan1}. 
The low field $M_L(\chi)$~\cite{Civale1, Kogan1} have sharp maxima around 
$[100]$ and $[010]$ directions, 
and broader minima around $[110]$ and $[1\overline{1}0]$ directions. 
The sharp maxima remind us of the observed cusp-like minima of 
thermal conductivity and specific heat~\cite{Izawa, Park1} 
around $[100]$ and $[010]$ directions. 
These have been argued as a result of so-called ``s+g'' 
pairing function~\cite{Izawa}. 
We calculated $M_L(\chi)$ for the pairing 
$w_{\bf p} \propto (1- \sin ^4 \theta \cos 4 (\phi+\chi) )$ with point nodes 
keeping $\beta=0.4$. 
However no such characteristic structure was observed 
in the angular dependence of $M_L(\chi)$. 
This means that explanations for the observed structure of $M_L(\chi)$ 
may need a different mechanism. 

In conclusion we have microscopically studied  
field-angle-resolved basal plane magnetization oscillations, 
and demonstrated not only 
that a careful measurement of the magnetization can be a 
potentially useful tool to identify the nodal position of the gap function 
when a material of interest possesses less anisotropic FS, 
but also that the experimental data for borocarbides 
are well reproduced by considering both gap and FS anisotropies. 
If combined with other field-angle-resolved quantities~\cite{Pedja2}, such as 
specific heat and thermal conductivity at low temperatures, we can further 
narrow down the possible pairing symmetry or gap anisotropy 
in various conventional and unconventional superconductors. 
The lesson from borocarbides tells us that 
for materials with {unusually} strong FS anisotropy such as borocarbides 
we should be careful to judge the nodal position through magnetization measurements, 
since {unusually} strong FS anisotropy can reverse the conclusion. 

After the submission of the first manuscript we learned about a preprint by 
Kusunose~\cite{Kusunose} which study the effect of the gap anisotropy alone 
based on a simplified version of our treatment. 
The result is consistent with ours when only gap anisotropy is considered. 

We would like to acknowledge useful discussions with T. Sakakibara and Y. Matsuda.

\bibliography{basename of .bib file}

\begin{thebibliography}{99}
\bibitem{motiv}
There is an ongoing controversy on the gap function; 
either $d_{x^2-y^2}$ or $d_{xy}$ 
pairings between angle-resolved thermal conductivity\cite{Izawa2} 
and specific heat\cite{Aoki} experiments in CeCoIn$_5$.
\end{thebibliography}

\end{document}